\documentclass[runningheads]{llncs}
\usepackage[T1]{fontenc}
\usepackage{graphicx}
\usepackage{amsmath}
\usepackage{bm}
\usepackage{esvect}
\usepackage[version=4]{mhchem}
\usepackage[utf8]{inputenc} 

\usepackage[table]{xcolor}

\graphicspath{./images/}

\usepackage[numbers]{natbib}

\begin{document}
\title{Thermodynamics of Biological Switches}
%
%
\author{Roger D. Jones\inst{1,2}\orcidID{0000-0002-8491-5421} \and
Achille Giacometti\inst{1}\orcidID{0000-0002-1245-9842} \and
Alan M. Jones\inst{2,3}\orcidID{0000-0002-2365-6462}}
\authorrunning{R. Jones, A. Giacometti, A. Jones}
%
\institute{European Centre for Living Technology,  
Department of Molecular Sciences and Nanosystems,
Ca’ Foscari, University of Venice, Italy \and
Department of Biology, University of North Carolina at Chapel Hill, USA 
\and
Department of Pharmacology, University of North Carolina at Chapel Hill, USA 
\email{RogerDJonesPhD@gmail.com}\\
}
\maketitle              
\begin{abstract}
We derive a formulation of the First Law of nonequilibrium thermodynamics for biological information-processing systems by partitioning entropy in the Second Law into microscopic and mesoscopic components and by assuming that natural selection promotes optimal information processing and transmission. The resulting framework demonstrates how mesoscopic information-based subsystems can attain non-equilibrium steady states (NESS) sustained by external energy and entropy fluxes, such as those generated by ATP/ADP imbalances {\it in vivo}. Moreover, mesoscopic systems may reach NESS before microscopic subsystems, leading to ordered structures in entropy flow analogous to eddies in a moving stream.

\keywords{biological switches \and nonequilibrium thermodynamics \and Second Law of Thermodynamics \and mutual information \and entropy \and free energy.}
\end{abstract}
\section{Introduction}

The universe is subject to the Second Law of Thermodynamics and is irreversibly moving toward heat death, a state of maximal disorder in which information and structure vanish \cite{reif2009fundamentals}. However, within this trajectory toward uniformity, life has emerged as a striking counterexample: an organized, information-processing phenomenon inseparably embedded in a turbulent physical environment. To persist, living systems must continually adapt, and adaptation depends on the acquisition, transmission, and transformation of information. In this sense, biology can be understood as a form of computation.

\begin{figure}
\includegraphics[width=\textwidth]{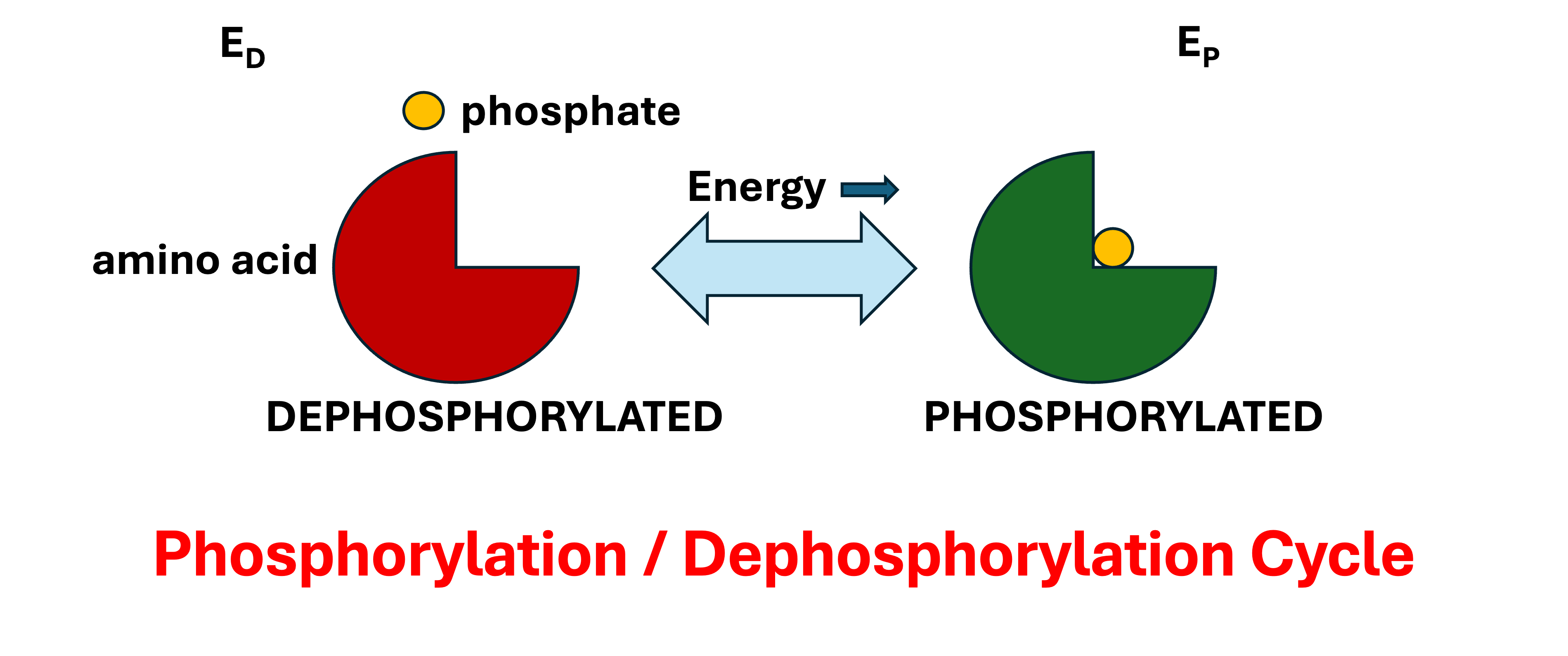}
\caption{An example of a simple chemical reaction commonly used as a biological switch.
Typically, energy input is required to phosphorylate an amino acid.
ATP imbalance drives the energy $E_D$ of the dephosphorylated state to the higher energy $E_P$ of the phosphorylated state.
The switch lives in a protein matrix that, through conformational changes, alter the energy relations between phosphorylated and dephosphorylated states.
Whether a site is phosphorylated or not can determine intracellular response to the switch state.} \label{PdPC}
\end{figure}

At the molecular scale, this computation is carried out by regulatory processes that function analogously to switches \cite{jones14model,jones2023proposed,jones2024information}. 
These mechanisms, fueled by external energy sources such as sunlight or nutrient-derived metabolites, govern cellular behavior through precise biochemical reactions. The phosphorylation/dephosphorylation cycle (PdPC) (see Figure \ref{PdPC}) exemplifies this mechanism \cite{qian2007phosphorylation}.
These reversible reactions, typically occurring in amino acids within proteins, act as decision points that direct cellular outcomes under a wide range of conditions. 
Other classes of molecular switches are widespread, yet share the defining characteristics of this cycle \cite{qian2007phosphorylation}.
Through their collective action, molecular switches dynamically regulate physiological functions, including immune responses, behavior, and metabolism. Life’s adaptability thus emerges from the orchestrated interplay of countless such reactions, which continuously compute and recalibrate biological information.

From a broader perspective, life represents a process by which order is carved from thermodynamic chaos. By coarse-graining innumerable microscopic states into stable functional patterns shaped by natural selection, biological systems maintain complexity despite entropic pressures. Sustained by external energy fluxes within an otherwise equilibrating universe, these systems manipulate information to create mesostates (nanometer length scale) that respond flexibly to environmental stimuli. In this way, life persists as a transient but resilient island of order, demonstrating how matter and energy can generate computational architectures capable of resisting the universal drift toward randomness.

Simple molecular switches, such as those illustrated in Figure \ref{PdPC}, lend themselves to analysis using statistical-physics methods similar to those described in the monograph by Seifert \cite{seifert2008stochastic}. 
Here, we apply statistical physics to a biological information-processing system immersed in a thermal bath and coupled to a protein matrix capable of modulating the switch’s energy levels. By incorporating Shannon’s concept of mutual information \cite{shannon1963mathematical} into the Second Law of Thermodynamics, we show that such systems can maintain functionality over extended timescales and operate at an extremum of information processing. This perspective further suggests a generalization of the First Law of Thermodynamics in which information itself contributes to the generation of heat.

This study complements a parallel investigation \cite{jones2025information} that derives the statistical ensemble for biological switches powered by dedicated energy sources. That framework, termed the Biological Ensemble, generalizes the Gibbs ensemble to describe open biological systems and shows that such switches can occupy quasistable states corresponding to on, off, and inactive configurations.

\section{The Model Switch}
\subsection{Microstates and Mesostates}

The model switch considered here, exemplified by the phosphorylation / dephosphorylation cycle (PdPC) shown in Figure \ref{PdPC}, consists of atoms and molecules whose precise configuration at any instant defines a microstate $x\in X$, where $X$ is the set of all possible microstates. The task of statistical mechanics is to determine the probability of a given microstate $x$ based on mesoscopic knowledge of the system and externally imposed constraints. In equilibrium statistical mechanics, this is accomplished by the Canonical Ensemble \cite{reif2009fundamentals}, from which the laws of equilibrium thermodynamics follow. In that case, the average energy of the microstates defines the temperature as a fundamental mesoscopic variable.

From a functional perspective, the switch is described by a mesostate $y$, where $y=(1,0)$ denotes the dephosphorylated form and $y=(0,1)$ denotes the phosphorylated form. Each mesostate comprises many underlying microstates; for example, numerous microscopic configurations are consistent with phosphorylation. Thus, the space of microstates can be partitioned according to their associated mesostates, and each mesostate $y$ is assigned a probability $\Pr(y)$. A rigorous treatment of these ideas is given in \cite{marletto2016constructor} and \cite{marletto2016constructorthermodynamics}.

The switch is embedded in a thermal bath of temperature $T$ and within a protein matrix whose configurations, determined by ligand binding, modulate the phosphorylation state. Both the bath and the protein matrix can themselves be described in terms of micro- and mesostates. Together, these components make up a mesoscopic information system that must be stable, programmable, and capable of transmitting information to other systems that generate biological outcomes. The heat bath mediates thermal exchange with this system.

Switches dissipate heat when they cycle through on and off states \cite{landauer1996physical}, and functional information processing requires the input of external energy. 
The fact that information generates heat upon its erasure is known as Landauer's Principle \cite{chattopadhyay2025landauer}.

For the PdPC, this energy is supplied by a nonequilibrium chemical flux $J$ arising from imbalances in ATP and ADP concentrations, though other nucleotides may play similar roles in different systems. Unlike equilibrium systems, biological information-processing systems are open and constrained by such fluxes, rendering the Canonical Ensemble of equilibrium thermodynamics inapplicable. Instead, the recently proposed Biological Ensemble \cite{jones2025information} generalizes the Canonical Ensemble to these open systems, yielding thermodynamic consequences distinct from those of equilibrium ensembles, including explicit treatment of heat dissipation from information processing \cite{landauer1996physical}.

\subsection{Natural Selection and Information}

We postulate that natural selection favors genetic architectures that enhance the efficiency of information processing and transformation. This efficiency improves the ability of organisms to adapt to fluctuating environments, thereby prolonging survival and increasing the persistence of genetic material across generations.
For single switch, this postulate leads to the observation that the switch can exist in up to three quasistable states for any specific values of the phosphate bonding energy \cite{jones2025information}.
This behavior has been observed in impedance measurements of photo-active ligands interacting with transmembrane receptors \cite{wirth2023monitoring}.

We adopt Shannon’s mutual information \cite{shannon1963mathematical} as a natural measure of biological information, since it follows from five fundamental axioms: non-negativity, symmetry, chain rule, data processing inequality, and additivity \cite{cover1999elements}. Relaxing one or more of these axioms produces alternative measures, but Shannon’s formulation remains the most widely applicable. For a mesoscopic system that transmits information from one storage state $y\in Y$ to another $y^{\prime}\in Y$, the transferred information $I_M$ is given by
\begin{equation}
I_M = \sum_{y\in Y} \sum_{y^{\prime}\in Y}
\Pr(y,y^{\prime}) \; \ln\left( \frac{\Pr(y,y^{\prime}) }{\Pr(y)\Pr(y^{\prime}) } \right)
\end{equation}
where $\Pr(y)$ and $\Pr(y^{\prime})$ denote the probabilities of the respective mesostates and $\Pr(y,y^{\prime})$ the joint probability distribution. For a switch such as the PdPC (Figure \ref{PdPC}), the set of mesostates can be written as
\begin{equation}
Y = { (d,p) }, \;\;\;\; p\in (0,1),\; d \in (0,1)
\end{equation}
where $d$ represents the dephosphorylated state and $p$ the phosphorylated state.

If natural selection acts to maximize information processing efficiency, the principle can be formalized as
\begin{equation}\label{maxInf}
d I_M \ge 0,
\end{equation}
implying that evolutionary dynamics preserve or increase the mutual information transmitted by biological switches.

\subsection{Second Law of Thermodynamics}

The path toward thermodynamics of biological systems proceeds through the Second Law of Thermodynamics, which holds in both equilibrium and nonequilibrium regimes. Adopting Jaynes’ informational formulation of entropy in the Second Law \cite{jaynes1957information}, we derive a corresponding First Law that describes the flow of energy among information, work, and the thermal bath.

The Gibbs-Shannon entropy $S$ of a classical physical system composed of a set $X$ of microstates $x \in X$ is (see \cite{jaynes1957information})

\begin{equation}\label{entropyAppendix}
    S = - \sum_{x\in X} \Pr(x) \ln \Pr(x)
\end{equation}
where $\Pr(x)$ is the probability that microstate $x$ occurs.

The Second Law of Thermodynamics can be stated as
\begin{equation}\label{secondLaw}
    dS \ge 0
\end{equation}
The fact that information processing generates heat implies that when the information of a mesosystem increases, the entropy of the microstates increases.

The entropy can then be restated as
\begin{equation}\label{information partition}
    S = S_m - I_M
\end{equation}
where $S_m$ is the entropy of microscopic states reduced by the degrees of freedom lost to the constraints on the information $I_M$. 
\begin{equation}\label{microEntropy}
    S_m =  \sum_{x\in X} \sum_{y\in Y} \sum_{y^{\prime}\in Y} 
    \Pr(x,y,y^{\prime})
     \; \ln\left(  \frac{       \Pr(y,y^{\prime})   }{\Pr(x) \Pr(y)  \Pr (y^{\prime} ) } \right)
\end{equation}
The minus sign in Equation \ref{information partition} is chosen in order to be consistent with Landauer's Principle.
Incorporation of finite information $I_M$ in Equation \ref{microEntropy} into microscopic entropy $S_m$ increases the entropy in microscopic states.
This is a consequence of the non-negativity of information flow $I_M$.

The Second Law becomes \cite{parrondo2015thermodynamics,seifert2008stochastic}
\begin{equation}\label{becoming}
    dS= dS_m - dI_M \ge 0
\end{equation}
and
\begin{equation}
    dS_m \ge dI_M
\end{equation}
which states that the heating of the microstates is faster or as fast as information is generated.
Equation \ref{microEntropy} defines the entropy for the microstates in the presence of controlled mesoscopic information $I_M$.

\subsection{Creation and Maintenance of Information}
Mesoscopic information $I_M$ is generated and maintained by externally driven chemical flux $J$ \cite{snoke2024crucial}. This flux can be expressed in terms of probability flow as
\begin{equation}\label{flux}
J = \Pr(y|y^{\prime}) \Pr(y^{\prime}) 
= \Pr(y^{\prime}|y)\Pr(y)
= \Pr(y,y^{\prime})
\end{equation}
where $\Pr(y|y^{\prime})$ and $\Pr(y^{\prime}|y)$ are the
transition probabilities between the mesostates $y$ and $y^{\prime}$.
In steady state, the forward flux from $y^{\prime}$ to $y$ is equal to the reverse flux from $y$ to $y^{\prime}$. For $y \neq y^{\prime}$, Equation \ref{flux} is Bayes theorem with $J = \Pr(y,y^{\prime})$ \cite{grover2012literature}.
This implies that the information can be completely specified with knowledge of $J$, $\Pr(y)$, and $\Pr(y^{\prime})$.
We can think of $J$ as an externally determined control parameter and take this quantity as given by external conditions.

It is important to note that Equation \ref{flux} does not imply microscopic reversibility of the mesosystem. The system need not follow the same trajectory in the transition from $y$ to $y^{\prime}$ as in the reverse transition from $y^{\prime}$ to $y$. The forward and reverse pathways may differ, forming a closed reaction cycle. This distinction is essential for biological switches such as PdPCs, in which the forward and reverse reactions are catalyzed by different enzymes, typically a kinase and a phosphatase, respectively. Such catalytic asymmetry enables enzymatic control over the state of the switch \cite{jones2025information}.

Since $I_M$ can be expressed in terms of ATP-driven flux $J$, it can be maximized by varying probabilities $\Pr(y)$ and $\Pr(y^{\prime})$ while keeping $J$ constant. This optimization yields quasistable states that correspond to the possible configurations of the switch \cite{jones2025information}. These quasistable states represent local maxima of information subject to the nonequilibrium steady-state (NESS) condition
\begin{equation}\label{NESS}
dI_M = 0.
\end{equation}
It follows from Equation \ref{becoming} that all entropy production occurs at the microscopic level,
\begin{equation}
dS = dS_m \geq 0,
\end{equation}
while the mesoscopic information states remain in NESS.

\section{First Law of Thermodynamics}
In the absence of work exchanged with the external environment, the First Law of Thermodynamics can be written using the Clausius relation between energy and entropy as 
\cite{reif2009fundamentals}
\begin{equation}
dU = T\; dS - p\;dV + dW
\end{equation}
where $U$ is the total energy of the system, $T$ is the constant temperature of the microstates, $p$ is the pressure, $V$ is the volume, and $dW$ is the work carried out on the system other than $p\; dV$ work.

Defining the Biological Free Energy $B$ as
\begin{equation}\label{BioFreeEnergy}
B = U + pV - TS_m = H - TS_m
\end{equation}
where $H$ is the enthalpy,
we obtain

\begin{equation}\label{newFirstLaw2}
dB = V\;dp\; -\; S_m\; dT\; -\; T\; dI_M\; +\; dW
\end{equation}
which, for constant temperature and pressure, becomes
\begin{equation}\label{newFirstLaw}
dB = -T\; dI_M + dW
\end{equation}
which shows that maximization of mesoscopic information $I_M$ in the absence of external work on the mesosystem corresponds to minimization of Biological Free Energy $B$. This condition is achieved when the mesoscopic system reaches a NESS ($dI_M = 0,\; dW=0$), which may occur well before the total entropy is maximized.
The timescales of entropy maximization in the universe are much longer than the timescales of natural selection.

The molecular switches reside within proteins capable of performing mechanical work
$dW$ on the switches through conformational changes in their tertiary structure. These conformational transitions are typically triggered by the binding and dissociation of ligands, coupling the chemical free energy to information processing within the protein. Consequently, the protein conformation determines or at least modulates the information content of the switch. The formation of a switch is influenced by the work generated as the protein matrix changes conformation, while the erasure of information is accompanied by the dissipation of heat into the surrounding microstates, consistent with Landauer’s Principle \cite{chattopadhyay2025landauer}. 
Thus, the creation and maintenance of mesoscopic information represent a dynamic partnership between the external flux
$J$, which drives energy through the mesosystem, and the mechanical work performed by the protein matrix on the information-bearing components. This framework unites thermodynamic, informational, and biological perspectives on cellular regulation.

Together, Equations \ref{microEntropy}, \ref{BioFreeEnergy}, and \ref{newFirstLaw} generalize the Gibbs free energy from equilibrium systems to biological switches operating in NESS.

\section{Discussion}

In biological information-processing systems, energy and entropy can be partitioned into microscopic and mesoscopic components. Introducing the concept of Biological Free Energy demonstrates that its minimization corresponds to the maximization of mesoscopic information. This condition defines a nonequilibrium steady state (NESS), which can be established well before total entropy is maximized. In this way, the framework generalizes the Gibbs free energy—traditionally restricted to equilibrium—to describe the dynamics of biological switches operating far from equilibrium.

Building on the separation of entropy in the Second Law, and assuming that natural selection favors optimal information processing and transmission, we derived expressions for the First Law of nonequilibrium thermodynamics in biological information systems (Equations \ref{BioFreeEnergy}, \ref{newFirstLaw}, and \ref{microEntropy}). The analysis assumes thermal equilibrium at the level of microstates, while mesoscopic states remain out of chemical equilibrium.

The mesoscopic information-processing subsystem reaches a NESS driven by external energy fluxes, such as ATP/ADP cycling. This steady state can emerge prior to microscopic equilibrium, generating structured entropy flows analogous to eddies in a moving fluid. On evolutionary timescales, life itself can be viewed as a persistent nonequilibrium phenomenon, sustained far from thermodynamic equilibrium long before universal entropy reaches its maximum.

It is important to emphasize that the present treatment does not require mesoscale information systems to be microscopically reversible, nor are the mesostates of biological switches assumed to be in or near thermal equilibrium, although the surrounding heat bath is considered equilibrated. The only requirement is that the switch operates in an NESS, where the information is located at an extremum.

The absence of microscopic reversibility implies that phosphorylation and dephosphorylation proceed along distinct reaction pathways, a necessity {\it in vivo}, where kinases catalyze phosphorylation and phosphatases drive dephosphorylation. Consequently, these reactions are thermodynamically coupled yet mechanistically distinct. The informational mesosystem thus functions far from thermochemical equilibrium. The probability distributions of switch states differ qualitatively from those of equilibrium systems; for a given phosphate-bonding energy, the system may occupy up to three quasistable states, in contrast to the single state permitted under equilibrium conditions \cite{jones2025information}.

The model of a biological switch regulated by a measurable chemical flux $J$ can be generalized to systems in which both total energy input and heat output are characterized by
$Q=Jq$,
where $Q$ represents the total rate of energy input from ATP hydrolysis and $q$ denotes the heat dissipated per switching cycle between on and off states. In networks of multiple switches, variations in heat dissipation among individual elements may determine the sequence of activation. Such thermodynamic coupling likely plays a critical role in switch regulation and will be explored further through combined theoretical and experimental studies.

Historically, stochastic thermodynamics \cite{parrondo2015thermodynamics,seifert2008stochastic} has addressed the resolution of Loschmidt’s paradox, which questions the compatibility of microscopic reversibility with the Second Law of Thermodynamics \cite{binder2023reversibility}. The time dependence of transient relaxation on a steady state, typically assumed to be equilibrium, has been central to this effort, leading to fluctuation theorems that describe the mesoscopic nature of these transients \cite{yang2023fluctuation}.

In contrast, the present treatment does not restrict the steady state to equilibrium; instead, it explicitly requires a NESS. Furthermore, we do not address the transient approach to NESS. The relevant notion of reversibility here is not microscopic, but logical reversibility \cite{chattopadhyay2025landauer}. If no work is performed on the mesosystem ($dW=0$) by the surrounding protein matrix, then NESS requires that no change occurs in the information content of the mesosystem ($dI_M=0$). This condition implies that the mesosystem is logically reversible, though not necessarily microscopically reversible.

The PdPC model developed here can easily be generalized to a wide range of biological switches, including networks of multiple interacting switches and information systems operating under diverse constraints. The entropy change of the microscopic states may differ across mesoscopic configurations, reflecting the dependence of molecular dynamics on higher-level organization. Transitions between on and off states can occur through conformational changes in the surrounding protein matrix, which functions as an external constraint shaping the accessible mesostates.

\begin{credits}

\subsubsection{\discintname}
The authors have no competing interests to declare
that are relevant to the content of this article. 
\end{credits}

\bibliographystyle{splncs04}
\bibliography{library}

\end{document}